\documentclass[pre,aps,twocolumn,nopacs,floatfix,superscriptaddress]{revtex4-1}

\usepackage{amsfonts,amssymb,amsmath,array}
\usepackage{graphicx}
\usepackage{amsbsy}
\usepackage{color}
\usepackage{hyperref} 
\usepackage{enumerate}
\usepackage{mathtools}
\usepackage{lipsum} 
\usepackage{cleveref}  

\begin{document}

\title{Inference of Time-Reversal Asymmetry from Time Series in a Piezoelectric Energy Harvester}

\author{L. Costanzo}
\affiliation{Department of Engineering, University of Campania ``Luigi Vanvitelli'', 81031~Aversa, Italy}

\author{A. Baldassarri}
\affiliation{Department of Physics, University of Rome Sapienza, P.le Aldo Moro 2, 00185, Rome, Italy}
\affiliation{Institute for Complex Systems - CNR, P.le Aldo Moro 2, 00185, Rome, Italy}

\author{A. Lo Schiavo}
\affiliation{Department of Engineering, University of Campania ``Luigi Vanvitelli'', 81031~Aversa, Italy}

\author{A. Sarracino}
\affiliation{Department of Engineering, University of Campania ``Luigi Vanvitelli'', 81031~Aversa, Italy}

\author{M. Vitelli}
\affiliation{Department of Engineering, University of Campania ``Luigi Vanvitelli'', 81031~Aversa, Italy}

\begin{abstract}We consider the problem of assessing the non-equilibrium
behavior of a system from the study of time series. In particular, we
analyze experimental data from a piezoelectric energy harvester driven
by broadband random vibrations where the extracted power and the
relative tip displacement can be simultaneously measured. We compute
autocorrelation and cross-correlation functions of these quantities in
order to investigate the system properties under time reversal. We
support our findings with numerical simulations of a linear
underdamped Langevin equation, which very well describes the dynamics
and fluctuations of the energy harvester. Our study shows that, due to
the linearity of the system, from the analysis of a single variable,
it is not possible to evidence the non-equilibrium nature of the
dynamics. On the other hand, when cross-correlations are considered,
the irreversible nature of the dynamics can be revealed.
\end{abstract}

\maketitle

\section{Introduction}

In this paper, we analyze experimental data from a
piezoelectric energy harvester~\cite{H3}, a~device that can convert
mechanical vibrations into electrical currents to feed small
sensors. When driven by broadband
vibrations~\cite{halv,costanzo2021stochastic}, energy harvesters
behave like Brownian motors~\cite{R02}, where random fluctuations
can be rectified to extract work. These systems have recently
attracted a lot of interest, both theoretical and experimental.
From the thermodynamic point of view, the~system extracts work from
a noisy environment. Does this violate the second principle of
thermodynamics, stated as Plank did: ``It is impossible to construct
a device which operates on a cycle and produces no other effect than
the production of work and the transfer of heat from a single
body''? The answer is obviously negative: the harvester converts
part of the energy extracted from mechanical noise to work and 
dissipates the rest in the form of heat (via mechanical friction and
Joule dissipation of the circuit resistance). Moreover, since the
harvester extracts a finite power, it is a non-equilibrium system,
despite its dynamics being stationary, i.e.,~invariant under time
translations.

The non-equilibrium nature of the system implies that its dynamics have
to be irreversible, i.e.,~it is not symmetric with respect to a time
inversion. Detecting and measuring such irreversibility can be
relevant for the study of the system, also in order to improve its
efficiency. However, the~determination of the time reversal asymmetry
can be a difficult task when only partial information is~available.

In fact, 
one of the central questions in the physical modeling of a system is
whether observations of a few variables can reveal the non-equilibrium
properties of the dynamics, namely, if it is reversible or not and if energy
currents are present~\cite{zwanzig2001nonequilibrium}. In~particular, in~experiments, one usually can only access a few
observables, and the answer to the above question can be not easy
without a full reconstruction of the phase
space~\cite{gnesotto2018broken,roldan2021quantifying}. From~the
theoretical perspective, it is also very interesting to address the
issue related to the estimation of relevant quantities, such as entropy
production, from~incomplete knowledge, especially for non-Markovian
systems~\cite{pigolotti2008coarse,puglisi2010entropy,teza2020exact}. Indeed,
in such cases, out-of-equilibrium features may appear even in the
absence of a net drift or a probability current.
The above issues have been generally addressed within the framework of
stochastic thermodynamics (ST)~\cite{seifertrev,BPRV08,temprev}, which
extends concepts of standard thermodynamics to the realm of stochastic
systems, with~applications in biology, e.g.,~molecular
motors~\cite{R02}, as~well as for micro-machines~\cite{Leonardo9541}
and other small systems~\cite{sientropy}. This approach has been
critically discussed by several authors~\cite{cohen2004note,gujrati2020jensen}.
An interesting question is related to the possibility of assessing the
equilibrium or non-equilibrium features of the system, analyzing a
time series of experimental
data~\cite{diks1995reversibility,crisanti2012nonequilibrium}. The~problem is notably relevant when the signal has Gaussian statistics, as~recently discussed in~\cite{lucente2022inference,lucente2023out}, in~particular in the one-dimensional~case.

Here we show that this is exactly the case for the
experimental system we are dealing with. The~piezoelectric energy
harvester is a device featuring a cantilever structure with a tip
mass, whose displacement in time $x(t)$ induces a voltage $v_p(t)$
across the electrical load. As~shown
in~\cite{costanzo2021stochastic,costanzo2022stochastic}, when driven
by broadband vibrations, this system can be very well modeled as a
generalized linear Langevin equation with an exponential memory
kernel, taking into account the electromechanical coupling between the
tip mass velocity and the voltage. Equivalently, the~same system can
be described by two coupled linear equations: an underdamped Langevin
equation describing the dynamics of the tip mass position $x(t)$ in a
parabolic potential and a deterministic linear equation for the
voltage $v_p(t)$. The~latter quantity can be easily measured in
experiments, while measuring the tip mass fluctuations requires a
laser to monitor its relative displacement. Here we consider
a time series of experimental data of both the voltage signal $v_p(t)$
and the voltage $v_z(t)$ measured by the displacement sensor probing
the tip mass motion. We show that from the computation of the
auto-correlation function of a single variable, it is not possible to
infer about the (non-)equilibrium nature of the system and the
irreversibility of its dynamics. On~the other hand, when the
appropriate cross-correlations between two different observables are
taken into account, the~time-asymmetry in the system can be~revealed.

The energy harvester and the related model considered in our paper
represent an instance of non-Markovian systems, where feedback
mechanisms and memory effects are at play. These kinds of models have
been the object of an intense study in recent years, from~the point of
view of stochastic
thermodynamics~\cite{sagawa2012nonequilibrium,munakata2014entropy,munakata2013feedback,rosinberg2015stochastic,doerries2021correlation,loos2019fokker,de2019oscillations,debiossac2022non,plati2023thermodynamic}.

The paper is structured as follows. In~Section~\ref{exp}, we describe
the experimental setup. In~Section~\ref{model}, we introduce the linear
model describing the piezoelectric energy harvester and in Section~\ref{discussion} we discuss the 
theoretical issues related to the assessment of
equilibrium/non-equilibrium features from time-series analyses. In~Section~\ref{data}, we fit the model parameters to the experimental
data and compute the appropriate correlation functions to reveal
time-reversal asymmetry. Finally, in~Section~\ref{conclusion}, we
discuss our findings and draw some~conclusions.

\section{Experimental~Setup}
\label{exp}

\begin{figure}[h!]
\includegraphics[width=0.45\textwidth]{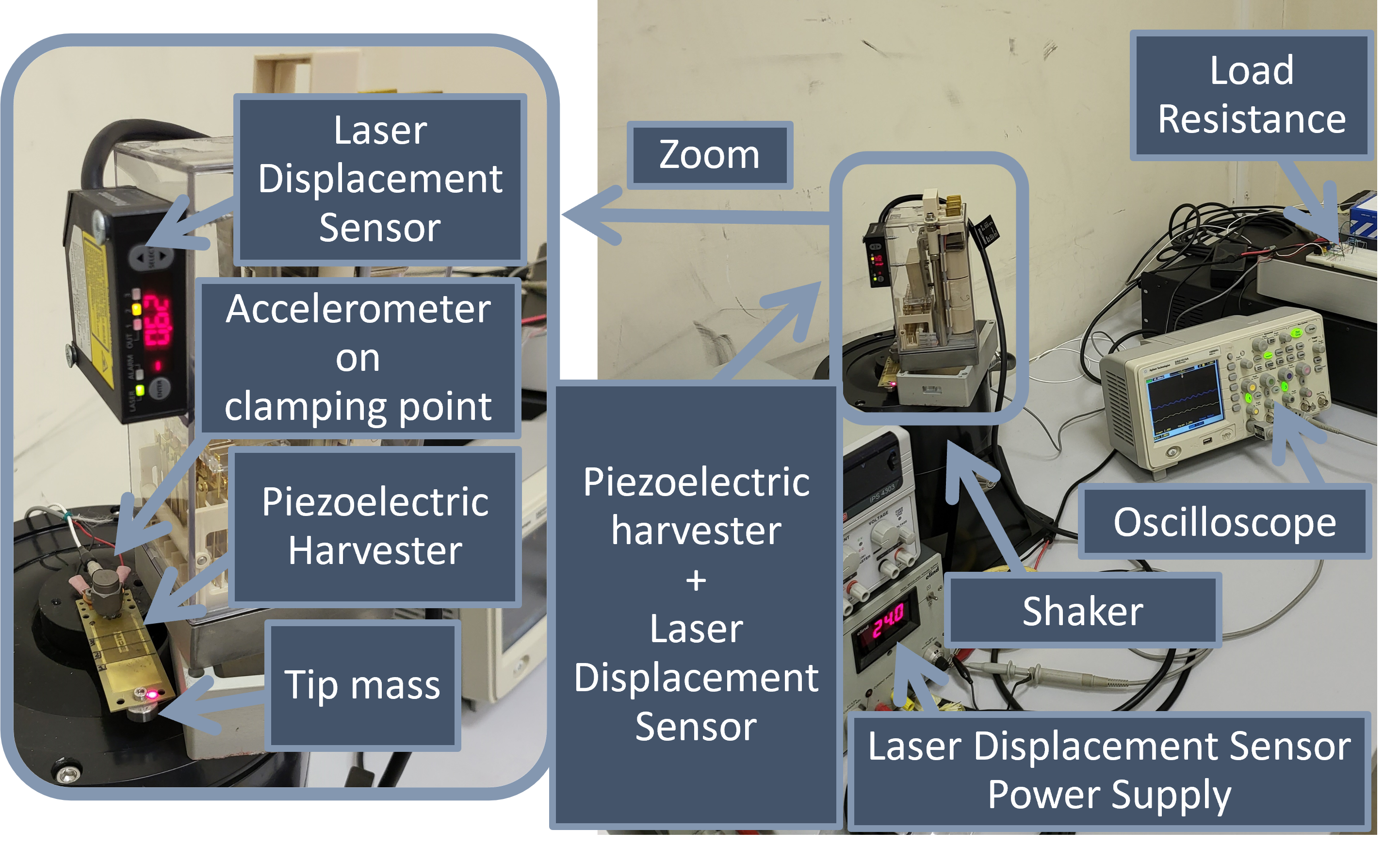}
\caption{Picture of the experimental apparatus.}
\label{fig_scheme}
\end{figure}

\begin{figure}[h!]
\includegraphics[width=0.45\textwidth]{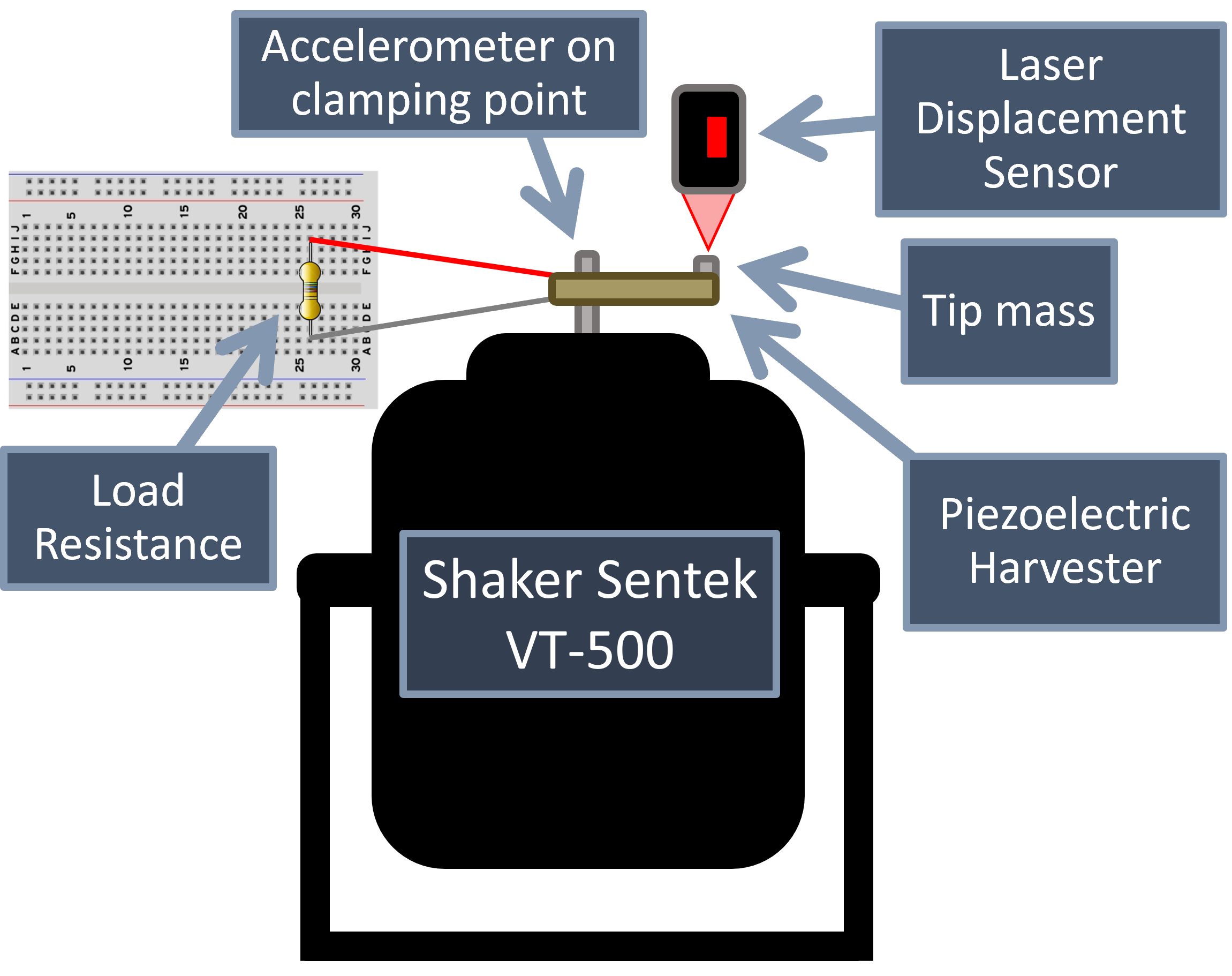}
\caption{Schematic representation of a cantilever structure with piezoelectric energy~harvester.}
\label{fig_scheme2}
\end{figure}

We consider a piezoelectric energy harvester with linear load. A~picture of the whole experimental apparatus is shown in
Figure~\ref{fig_scheme}, and its detailed description can be found
in~\cite{costanzo2021stochastic}. A~schematic representation of the
experimental setup for the piezoelectric harvester is shown in
Figure~\ref{fig_scheme2}. It consists of a cantilever structure with a
tip mass, which is vibrated by an electrodynamic shaker Sentek VT-500  (Sentek Dynamics, Santa Clara, CA, USA) 
inducing a voltage $v_p(t)$ across an electrical load resistance
$R$. We used the commercial piezoelectric harvester MIDE PPA-4011 (Mide Technology Corporation, MA, USA)
loaded by different electrical load resistances. The~harvester was
mounted on the shaker head in correspondence with a clamping point where
the accelerometer Dytran 3055D2 (Dytran Instruments, Chatsworth, CA, USA) 
was placed for measuring the input
acceleration $a(t) = \ddot{y}(t)$. A~small mass was added on the tip
of the harvester, and its displacement was monitored by the Laser
Displacement Sensor Panasonic HL-G112-A-C5 (Panasonic Corporation, Tokyo, Japan). 
The movement
of the tip mass occurs only along the vertical direction. In this
setup, the~two quantities that can be directly accessible are the
voltage $v_p(t)$ across the load resistance $R$ and the voltage
$v_z(t)$ measured by the sensor and proportional to the~displacement.

To study the response of the system to broadband vibrations, we fed
the shaker with a Gaussian signal generated with standard software
(MATLAB, R2023b) 
with~a sampling rate $f=5$ kHz. From~the measurement of the
voltage $v_p$, we obtain the average extracted power $P_{harv}=\langle
v_p^2\rangle/R$, where $R$ is the load resistance and $\langle \ldots
\rangle$ denotes an average in the stationary state. The~study of the system for different vibration amplitudes and for a wide
range of load resistances is reported
in~\cite{costanzo2021stochastic}.

\section{Theoretical~Model}
\label{model}

The piezoelectric energy harvester introduced in the previous section
can be accurately described with the
following linear model~\cite{costanzo2021stochastic}:
\begin{equation}
\begin{cases}
\dot{x}=v, \label{lang1}\\
M\dot{v}=-K_sx -\gamma v -\theta v_p + M \xi, \\
C_p\dot{v}_p=\theta v -\frac{v_p}{R}, 
\end{cases}
\end{equation}
where $\xi$ is white noise with zero mean and correlation $\langle
\xi(t)\xi(t')\rangle = 2D_0\delta(t-t')$. In~the above equations, $x$
represents the displacement of the effective tip mass $M$ {  along the vertical direction}, $v$ its
velocity, $\gamma$ the viscous damping due to the air friction, $K_s$
the stiffness of the cantilever in the harmonic approximation, $v_p$
the voltage across the load resistance $R$, $C_p$ the effective output
capacitance of the piezoelectric transducer, and~$\theta$ its effective
electromechanical coupling~factor.

As already observed in~\cite{costanzo2021stochastic}, the~system of
equations can be rewritten as a non-Markovian model, where a memory is present in the dynamics. In~particular, one has
\begin{equation}
\dot{v}=-\int_0^t\left[2\frac{\gamma}{M}\delta(t')+\Gamma(t-t')\right]v(t')dt'-\frac{K_s}{M}x+\xi,
\end{equation}
where the memory kernel $\Gamma(t)$ has the exponential form
\begin{equation}
\Gamma(t)=\frac{\theta^2}{C_pM}e^{-t/RC_P}.
\end{equation}

From 
this formulation of the model, one can immediately observe that
the friction memory kernel $\Gamma(t)$ is not associated with any
noise term. This makes the system out of equilibrium, because~the
fluctuation--dissipation relation of the second kind is explicitly
broken. Note that a noise term on the dynamical equation of the
voltage could be also considered, due to the presence of thermal
fluctuations in the circuit (Johnson noise). However, in~this setup,
such a term is negligible. In~order to make it relevant, the~  amplitude of the shaker should be significantly~reduced.

\subsection*{Stationary~Quantities}

For the linear model introduced above the static properties can be easily obtained by standard~methods~\cite{R89}.
Introducing the column vector 
${\bf X}=(x,v,v_p)^T$ and the coupling~matrix
\begin{equation}
{\bf A}=\left(
\begin{array}{ccc}
0 & -1 & 0 \\
\frac{K_s}{M} & \frac{\gamma }{M} & \frac{\theta }{M} \\
0 & -\frac{\theta }{C_p} & \frac{1}{C_p R}
\end{array}
\right),
\end{equation}
Equation~(\ref{lang1}) can be rewritten in vectorial form as
\begin{equation}
\dot{{\bf X}}=-{\bf A} {\bf X} + {\boldsymbol \eta},
\end{equation}
where ${\boldsymbol \eta}=(0,\xi,0)^T$.
Introducing the covariance matrix ${\boldsymbol \sigma}=\langle X^T X \rangle$ as
\begin{equation}
{\boldsymbol \sigma}=\left(
\begin{array}{ccc}
\sigma_{xx} & \sigma_{xv} & \sigma_{xv_p} \\
\sigma_{vx} & \sigma_{vv} & \sigma_{vv_p} \\
\sigma_{v_px} & \sigma_{v_pv} & \sigma_{v_pv_p}
\end{array}
\right),
\end{equation}
at stationarity, one has the constraint
\begin{equation}\label{stationary}  
{\boldsymbol D}=\frac{{\bf A}{\boldsymbol \sigma}+{\boldsymbol \sigma} {\bf A}^T}{2},
\end{equation}
where ${\bf D}$ is the noise matrix
\begin{equation}
{\bf D}=\left(
\begin{array}{ccc}
0 & 0 & 0 \\
0 & D_0 & 0 \\
0 & 0 & 0
\end{array}
\right).
\end{equation}

The 
stationary distribution is a multivariate Gaussian
\begin{equation}  
P_{st}(x,v,v_p)\sim e^{-\frac{1}{2}\left(\sigma_{xx}^{-1}x^2+\sigma_{vv}^{-1}v^2+\sigma_{v_pv_p}^{-1}v_p^2+2\sigma_{xv}^{-1}xv+2\sigma_{xv_p}^{-1}xv_p+2\sigma_{vv_p}^{-1}vv_p\right)},
\end{equation}
where ${\boldsymbol \sigma}^{-1}$ denotes the inverse matrix of ${\boldsymbol \sigma}$. 
From this expression, we can obtain the marginalized probability distribution of the quantity $v_p$ as
\begin{equation}  
P(v_p)=\int_{-\infty}^{\infty}dx \int_{-\infty}^{\infty}dv P_{st}(x,v,v_p) =\frac{1}{\sqrt{2\pi}\sigma_{v_p v_p}} e^{-\frac{v_p^2}{2\sigma^2_{v_pv_p}}},
\end{equation}
where
\begin{equation}  
\sigma_{v_pv_p}=\langle v_p^2\rangle=\frac{D_0 \theta^2 M^2 R^2}{\gamma  C_p R \left(\gamma +C_p K_s R+\theta^2 R\right)+M \left(\gamma +\theta^2 R\right)}. 
\end{equation}

The average output power is then given by
\begin{equation}\label{pharv0}  
P_{harv}=\frac{1}{R}\langle v_p^2\rangle.
\end{equation}

\begin{figure}[ht!]
\includegraphics[width=0.45\textwidth]{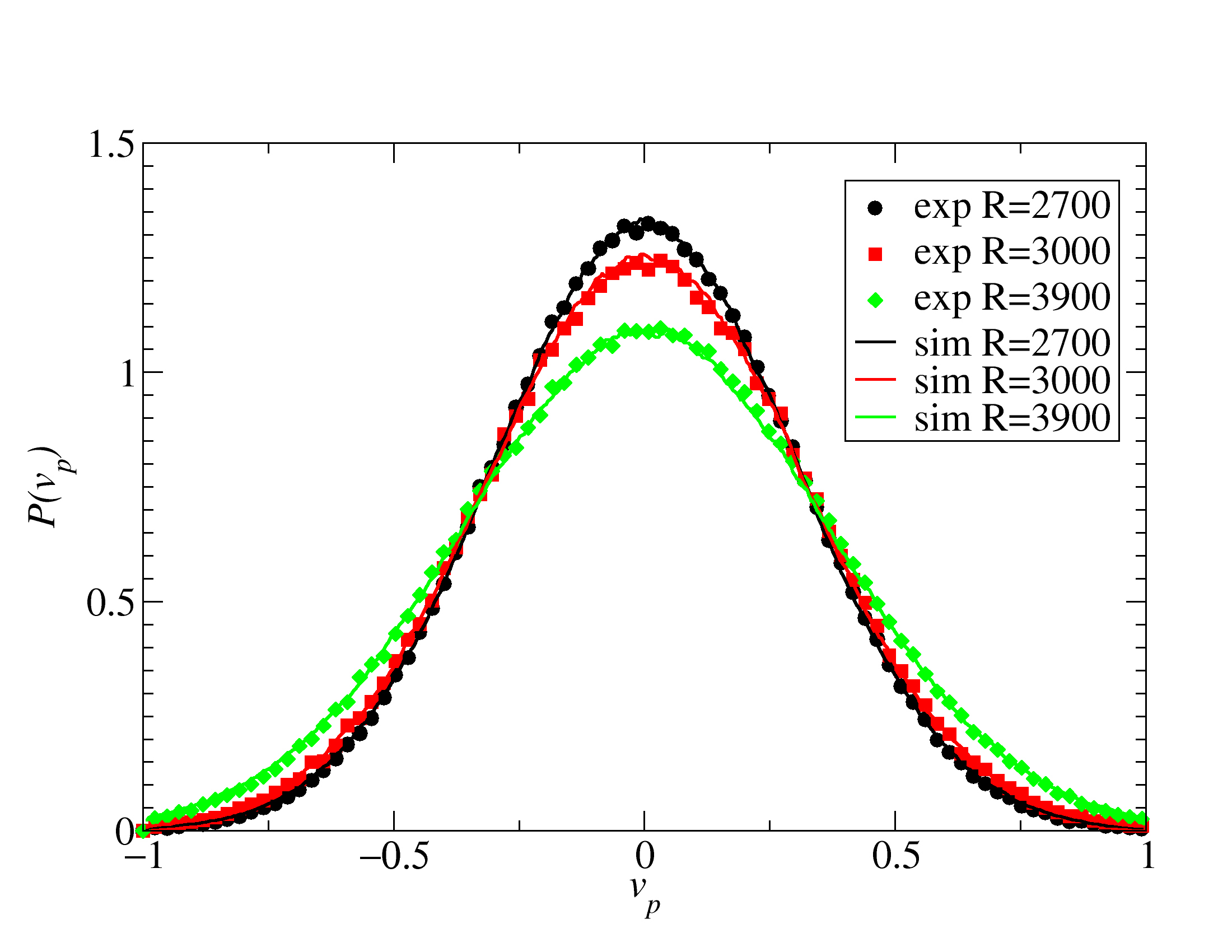}
\caption{Comparison 
between the experimental data (symbols) and the
theoretical model (lines). The~probability distributions $P(v_p)$
measured in experiments and in numerical simulations show a very
good agreement, with~the parameters reported in the main text
and load resistances \mbox{$R\in [2700, 3900]$ $\Omega$}.}
\label{fig0}
\end{figure}

The linear model described by Equations~(\ref{lang1})
features physical parameters that can be directly controlled in the
experiments and others whose values can be fitted to match the
measured data. The~amplitude of the white noise $D_0$ is related to
the acceleration $a$ provided by the shaker and to the sampling rate
$1/\Delta t$ of the input signal, $D_0=a^2\Delta t/2$, where $\Delta
t=1/f=0.0002$ s. The~mechanical parameters $K_s$ and $M$ are related
to the characteristic frequency of the system, which for the
experimental apparatus is $\sqrt{K_s/M}=2\pi\times 91$ Hz. Finally,
the capacitor $C_p$ is estimated as $C_p\sim 490$ nF. The~other
parameters can be fitted to the experimental data through
Equation~(\ref{pharv0}) as a function of $R$. We performed several
experiments with different values of $R\in [2700-3900]$ $\Omega$ and
measured the average extracted power $P_{harv}$. From~the fit, we
obtain the following values: $M=0.0198\pm 0.005$ Kg,
$\theta=0.00895\pm 0.0005$ N/V, $\gamma=0.3592 \pm 0.005$ Kg/s. The~shaker acceleration is $a=0.8\times 9.81$ m/s$^2$. Using these values of
the parameters, we performed numerical simulations of the model in
Equation~(\ref{lang1}) to measure the quantity $v_p(t)$ in the
case $R=2700,3000,3900$ $\Omega$ and its distributions. The~numerical
results are in very good agreement with experimental data (see
Figure~\ref{fig0}), showing that the linear model very well describes the
dynamics of the system. Once assessed the effectiveness of the
theoretical model, we next focus on the study of the non-equilibrium
properties of the system, as~discussed in the following~section.

\section{Inference of Time-Reversal~Asymmetry}
\label{discussion}

The non-equilibrium nature of a system is reflected in the
time-reversal asymmetry of the dynamics, which can manifest in
different quantities and can be checked in several ways (see, for
instance, the accurate discussion reported
in~\cite{lucente2022inference} and the
review~\cite{zanin2021algorithmic}). The~more common
quantity introduced to study the off-equilibrium character of a
system is the fluctuating entropy production (EP), which can be
defined at the level of a single trajectory of the dynamics,
considering the ratio between the probability of observing a
trajectory in the phase space and the probability of observing the
time-reversed trajectory (obtained by visiting the same states in
the phase space, but~in reversed order)~\cite{LS99}. This quantity
strictly vanishes in equilibrium, where detailed balance holds, but~is different from zero when energy currents are present and can be
considered as a measure of how far the system departs from
equilibrium. To~measure such a quantity requires direct access to
all degrees of freedom in the system that are responsible for
non-equilibrium behaviors. In~numerical simulations, this is
generally possible, whereas in real experiments the task is usually
much more difficult. Indeed, often, one can obtain only partial
information on the system under study due to the limitation of the
experimental apparatus that can measure some specific quantities.
Therefore, in~recent years, a lot of attempts have been made to provide
estimations or bounds on EP from the measurement of currents directly
accessible in experiments. This led to the development of the
so-called thermodynamic uncertainty relations, opening a large field
of
research~\cite{PhysRevLett.114.158101,roldan2021quantifying,manikandan2020inferring,ghosal2022inferring,martinez2019inferring,uhl2018fluctuations,kim2022estimating,dechant2023thermodynamic,dechant2023thermodynamic2}.

Other approaches to assess non-equilibrium behaviors are based on the
application of the fluctuation--dissipation relation. For~systems in thermal equilibrium, this relation takes a very simple
structure, relating the response of a quantity to an external
perturbation with the fluctuations of the same quantity (namely the
auto-correlation function) computed in the unperturbed dynamics, the~  constant of proportionality between response and autocorrelation
being the temperature. On~the other hand, in~non-equilibrium
conditions, violations of this relation can be observed, revealing
the off-equilibrium dynamics of the
system.~\cite{K66,R98,ss06,BMW09,sarra10b,S13,GPSV14,temprev}. However,
the study of the validity of the equilibrium relation requires performing response experiments. This means that a perturbation of the
system, changing some external parameters, has to be introduced in
order to measure the response function. This is not always simple in
real experiments. Let us also mention that generalized
Fluctuation-Dissipation Relations, valid also out of equilibrium
have been derived~\cite{BPRV08}, but~these take a more complex
structure and involve cross-correlations between different degrees
of~freedom.

Another interesting way to put in evidence the time-asymmetric
character of the dynamics is based on the analysis of the shape of the
avalanches reconstructed from the signal. An~avalanche is defined as a
region of the stochastic trajectory between two successive passages
through a given threshold value. This analysis plays a central role in
different physical systems, see for
instance~\cite{blumenthal2012excursions,majumdar2015effective,baldassarri2021universal,de2021critical},
where asymmetric shapes of the avalanche profiles can be observed.
The difficulty of such an analysis depends on the fact
that a large amount of statistics is necessary to accurately reconstruct the avalanche
profiles.

Here we will address the problem by exploiting a further approach, which
focuses on the time-reversal properties of higher-order correlation
functions. This alternative way to reveal irreversible dynamics was
first proposed in~\cite{pomeau1982symetrie}, and~has been recently
applied for granular systems in~\cite{lucente2023out}. 
See also Ref.~\cite{pomeau2017langevin} for an interesting
discussion on correlations in the Langevin equation. In~general, it
can be shown that a system is at equilibrium if the correlation
$C_{fg}(t)\equiv\langle f(t)g(0)\rangle=\langle
g(t)f(0)\rangle=C_{gf}(t)$, for~all functions $f$ and $g$. For~stationary systems, this is equivalent to the condition
$C_{fg}(t)=C_{fg}(-t)$. However, when the system is linear, the~time-reversal asymmetry cannot be revealed by the study of a single
observable, and cross-correlations have to be taken into account. The~main aim of this work is to provide an explicit example of this
problem with the analysis of experimental data, as~discussed in the
following.

\subsection{Data~Analysis}
\label{data}

\begin{figure}[h!]
\includegraphics[width=0.45\textwidth]{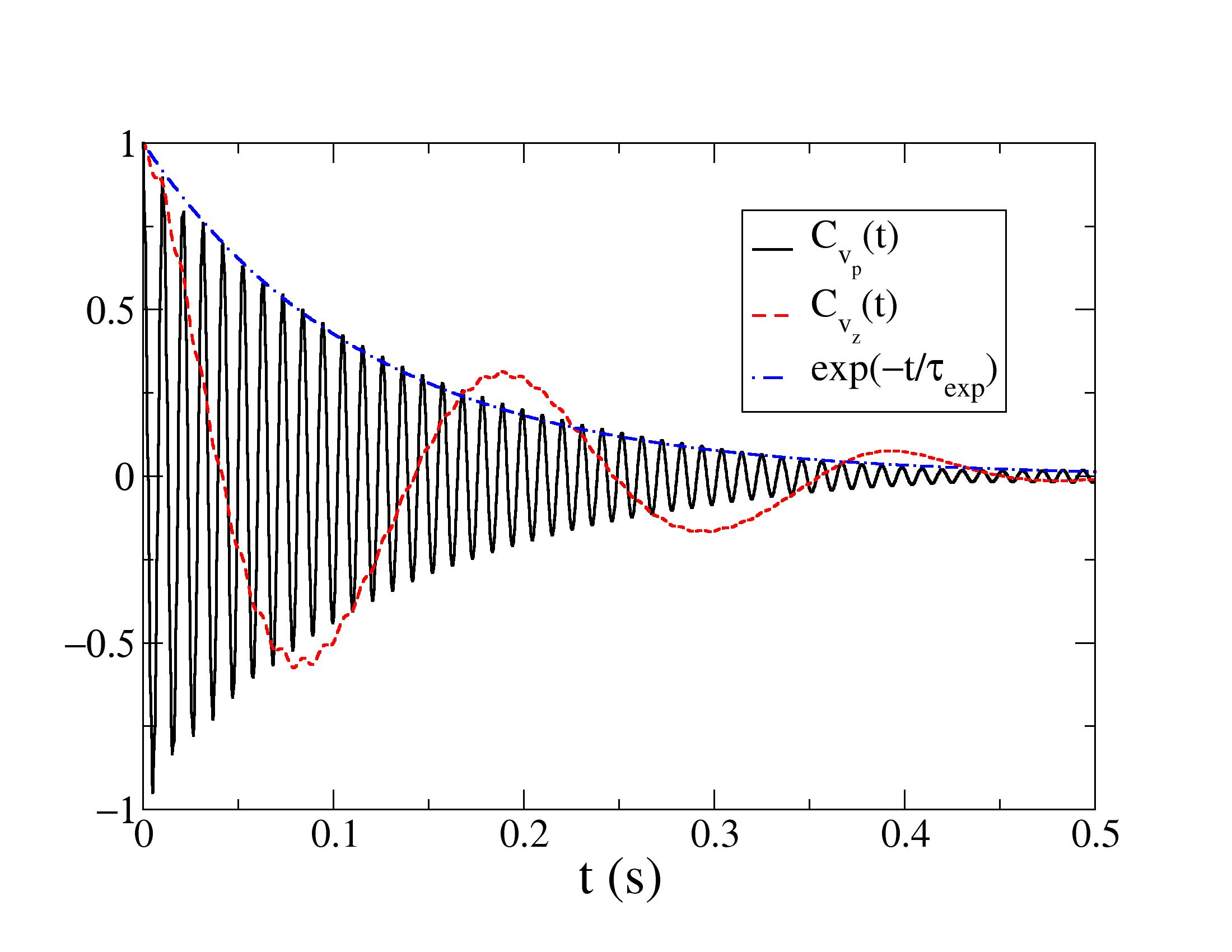}
\caption{Temporal 
autocorrelation functions of $v_p(t)$ and $v_z(t)$ measured in
experiments.}
\label{fig1a}
\end{figure}

We address the issue related to the (non-)equilibrium nature of the
system and its time-reversal symmetries analyzing time series of two
measured quantities: $v_p(t)$, the~voltage at the resistance load, and~$v_z(t)$, the~voltage at the displacement sensor, related to the tip
mass dynamics. We first show in Figure~\ref{fig1a} the normalized temporal
autocorrelation functions of $v_p(t)$ and $v_z(t)$,
\begin{eqnarray}\label{corr0}  
C_{v_p}(t)&\equiv& \frac{\langle v_p(t)v_p(0)\rangle}{\langle v_p(0)v_p(0)\rangle}, \\
C_{v_z}(t)&\equiv& \frac{\langle v_z(t)v_z(0)\rangle}{\langle v_z(0)v_z(0)\rangle},
\end{eqnarray}
that are characterized by an exponential decay, modulated by an
oscillatory behavior. The~characteristic decay time is estimated as
$\tau_{exp}\simeq 0.118\pm 0.005$ s, while the oscillation period is
$T_{exp}\simeq 0.010\pm 0.005$ s. These values are in good agreement
with the theoretical values obtained from the study of the eigenvalues
of the matrix ${\bf A}$ of the model, which give $\tau=0.079$ s and $T=0.0109$
s. An~accurate study of the analytical form of time correlation functions in non-Markovian systems
is presented in~\cite{doerries2021correlation}.

By construction, from~these two-point autocorrelation functions, one
cannot extract any information on the irreversible dynamics of the
system. In~order to define correlation functions with a single
variable that could unveil a time-asymmetry, one has to consider
higher-order correlators. In~particular, from~the time series of
$v_p(t)$ and $v_z(t)$, we compute the following normalized correlation
functions:
\begin{eqnarray}\label{corr}  
C^{(4)}_{v_p}(t)&\equiv& \frac{\langle v_p(t)v^3_p(0)\rangle}{\langle v^4_p(0)\rangle}, \\
C^{(4)}_{v_z}(t)&\equiv& \frac{\langle v_z(t)v^3_z(0)\rangle}{\langle v^4_z(0)\rangle}, \\
\end{eqnarray}

We 
note that, since the averages $\langle v_z(t)\rangle=\langle
v_p(t)\rangle=0$, the~total order of the correlation must be even
(otherwise the normalization at zero time would be zero), and~therefore, we consider the correlation between a variable and its
third~power.

\begin{figure}[h!]
\includegraphics[width=0.45\textwidth]{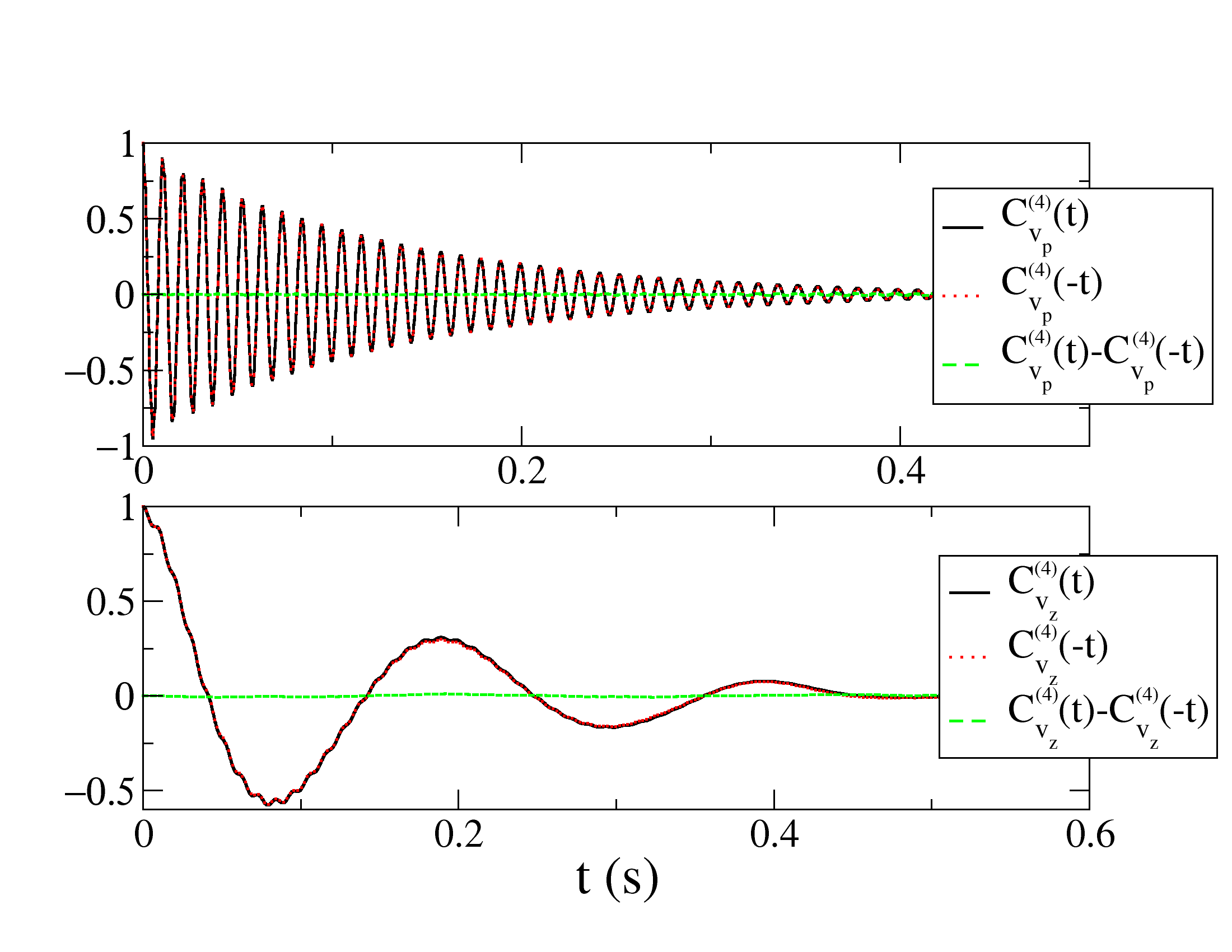}
\caption{High-order 
correlation functions $C^{(4)}_{v_p}(t)$ and
$C^{(4)}_{v_z}(t)$ {obtained from real data} measured in experiments. We note that from the
analysis of a single variable the dynamics appears symmetric
under time-reversal, as~evidenced by $C^{(4)}_{v_z}(t) - C^{(4)}_{v_z}(-t)$,
which is zero at all times.}
\label{fig1}
\end{figure}

For a linear system, it has been shown~\cite{lucente2022inference}
that from the observation of a single quantity, one cannot obtain
information on the non-equilibrium nature of the system.
This is due to the fact that linear systems present
Gaussian statistics. Therefore, a multi-point correlator can be
expressed in terms of two-point correlation functions, which are
always symmetric under time-reversal by construction. Our
experimental data illustrate this mechanism. Indeed, as~reported in
Figure~\ref{fig1}, from~the analysis of a single observable, even
considering higher order correlation functions, time-reversal symmetry
cannot be observed. This is clear from the behavior of the functions
$C^{(4)}_{v_p}(t)-C^{(4)}_{v_p}(-t)$ and
$C^{(4)}_{v_z}(t)-C^{(4)}_{v_z}(-t)$, which vanish in both cases.
This result further corroborates the efficacy of the linear model
Equations~(\ref{lang1}) to describe the experimental setup: indeed, the~presence of some relevant nonlinearities in the system could make the
correlation functions $C^{(4)}_{v_p}(t)$ and $C^{(4)}_{v_p}(-t)$
different due to the out-of-equilibrium nature of the~system.

\begin{figure}[h!]
\includegraphics[width=0.45\textwidth]{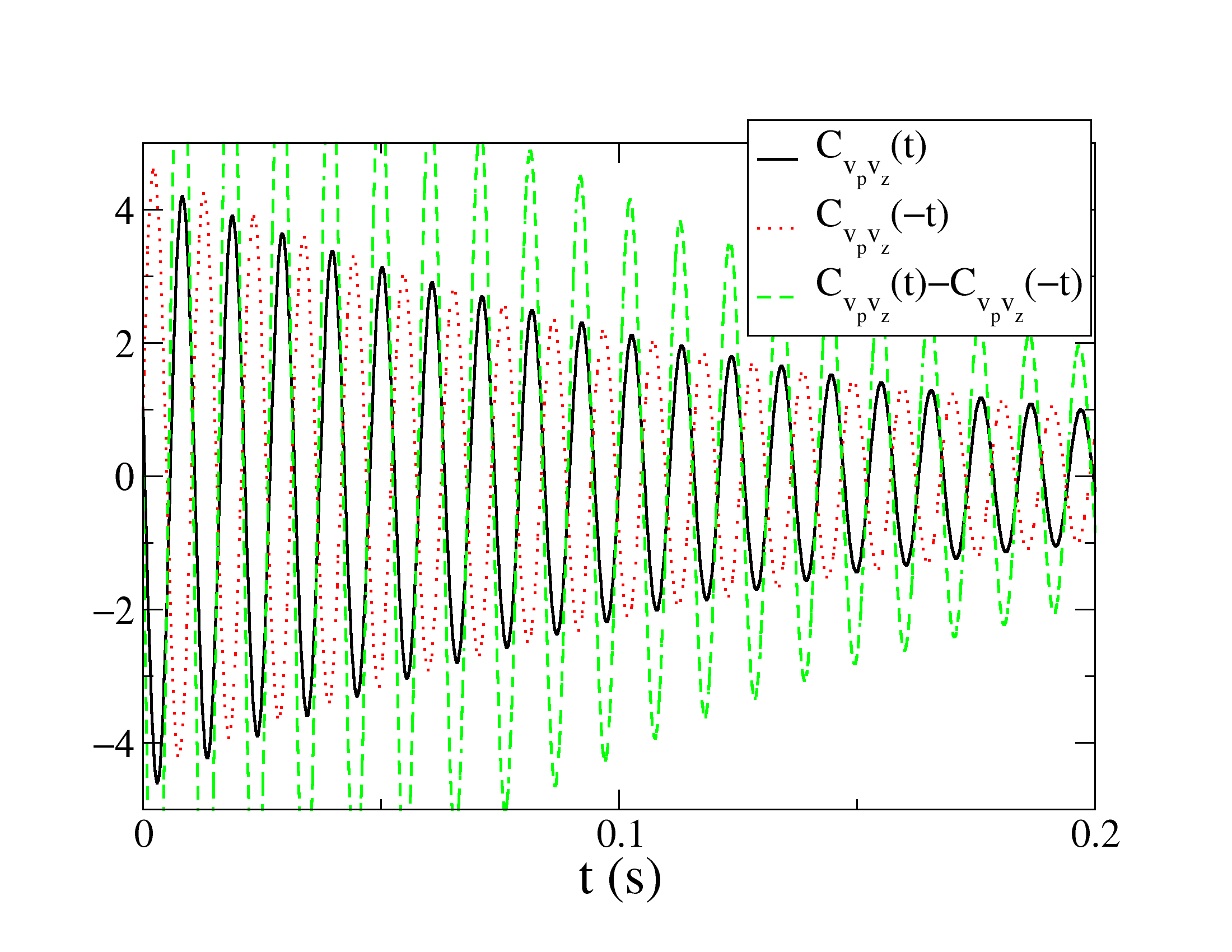}
\caption{Cross-correlation 
function $C_{v_pv_z}(t)$  obtained from real data measured in
experiments. The~non-equilibrium time-asymmetric dynamics of the
system clearly appears from the observation of the coupled motion of
$v_p$ and $v_z$, as~evidenced by $C_{v_pv_z}(t) - C_{v_pv_z}(-t)$,
which differs from zero.}
\label{fig1b}
\end{figure}

On the other hand, the~time-reversal
asymmetry in the model can be clearly revealed
if one considers the cross-correlation between the two coupled
variables $v_p(t)$ and $v_z(t)$
\begin{eqnarray}\label{corr1}  
C_{v_pv_z}(t)&\equiv& \frac{\langle v_p(t)v_z(0)\rangle}{\langle v_p(0)v_z(0)\rangle}, \\
C_{v_zv_p}(t)&\equiv& \frac{\langle v_z(t)v_p(0)\rangle}{\langle v_p(0)v_z(0)\rangle}= C_{v_pv_z}(-t),
\end{eqnarray}
as shown in Figure~\ref{fig1b}.

\subsection{Numerical~Simulations}

\begin{figure}[h!]
\includegraphics[width=0.45\textwidth]{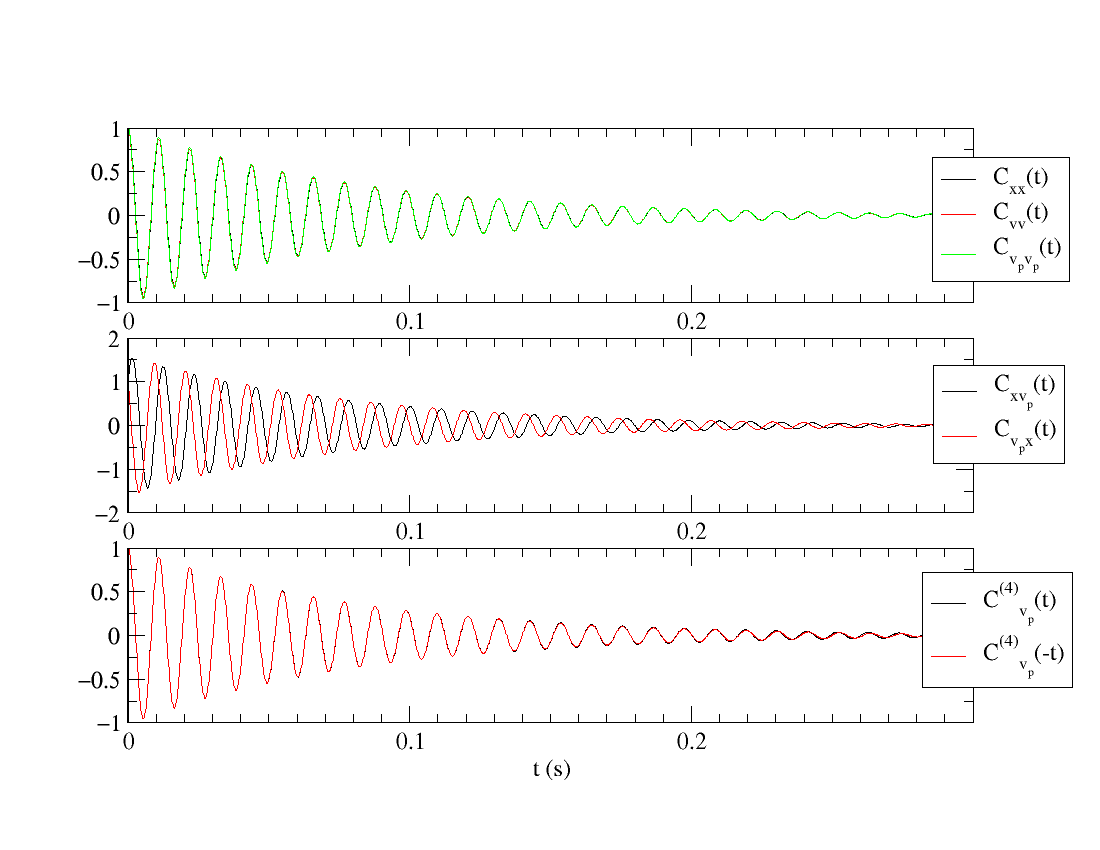}
\caption{Correlation functions obtained in numerical simulations
  (parameter values are the same as those fitted to the experimental
  data with $R=3000$ $\Omega$). In~the top panel, the~autocorrelations
  of position, velocity and load voltage are reported, showing all the
  same behavior. The~cross correlations between $x$ and $v_p$ shown in
  the middle panel reveal the time-asymmetry of the dynamics. In~the
  bottom panel, high-order correlations of $v_p$ are reported: both
  $C^{4}_{v_p}(t)$ and its time-reversed $C^{4}_{v_p}(-t)$ show the
  same~behavior.}
\label{fig2}
\end{figure}

To further support our findings, we performed the same analysis on
the numerical simulations of the model Equation~(\ref{lang1}).
In particular, we numerically computed the voltage $v_p(t)$, the~velocity $v(t)$, and the position $x(t)$, which can be directly accessed
in the simulations. The~results reported in Figure~\ref{fig2} confirm
the scenario described above for the experimental data: analyses of
(high-order) correlations of a single variable do not allow one to
reveal a non-equilibrium behavior (top and bottom panels), whereas
cross-correlation functions show the breaking of the time-reversal
symmetry (middle panel).

\section{Conclusions}
\label{conclusion}

We presented an experimental and theoretical study of a
piezoelectric energy harvester driven by white noise vibrations, with~the aim to illustrate the problem of extracting information on
non-equilibrium dynamics of a system from real time series. We have
first shown that the system can be well described by an effective
linear model, featuring an inertial Brownian particle in a parabolic
potential, coupled with an auxiliary variable. This second quantity
describes the electromechanical coupling between the tip mass of the
piezoelectric device with the electronic circuit. The~total system is
by construction out of equilibrium because the auxiliary variable is
not in contact with any~thermostat.

We then focused on the analysis of the temporal correlation functions
of the measured quantities, load voltage $v_p(t)$ and displacement
voltage $v_z(t)$, in~order to illustrate the problem of assessing the
non-equilibrium behavior of the system. As~expected, from~the
observation of a single quantity, even for higher order correlations,
the time-reversal asymmetry cannot be revealed due to the linearity
of the model. However, when cross-correlations are considered, the~non-equilibrium, time-asymmetric dynamics are unveiled. Numerical
simulations of the theoretical model confirm the emerging~scenario.

This work paves the way for other studies concerning non-equilibrium
features, such as entropy production, in~this kind of experimental
system, where memory and feedback effects are
present. {For instance, the~study of the entropy
	production can provide information on the distance from equilibrium
	of the system as a function of the model parameters.}  In
particular, the~relevance of fluctuations in our setup can be
described via the introduction of the concept of effective
temperature~\cite{temprev}, related to the amplitude vibration of the
shaker. This approach is common in macroscopic athermal systems, such
as vibro-fluidized granular fluids~\cite{puglisi13}, where the notion
of effective temperature can be useful to try to extend the
theoretical framework of statistical mechanics to non-equilibrium
systems. In~our case, a~similar treatment could be attempted, and a
definition of entropy production could be provided in terms of such an
effective parameter. We plan to study this generalization in our model
in future~works.

\vspace{6pt} 

\acknowledgments{A.S. and A.B. thank A. Puglisi and A. Vulpiani for useful~discussions.}


\section*{References}

\end{document}